\documentclass[lineno]{biometrika}
\usepackage{amsmath,mathtools,comment, times, bm, natbib,caption}
\usepackage{color}
\usepackage[plain,noend]{algorithm2e}
\usepackage{url}

\makeatletter
\renewcommand{\algocf@captiontext}[2]{#1\algocf@typo. \AlCapFnt{}#2} 
\def\@algocf@capt@plain{top}
\renewcommand{\algocf@makecaption}[2]{%
  \addtolength{\hsize}{\algomargin}%
  \sbox\@tempboxa{\algocf@captiontext{#1}{#2}}%
  \ifdim\wd\@tempboxa >\hsize
    \hskip .5\algomargin%
    \parbox[t]{\hsize}{\algocf@captiontext{#1}{#2}}
  \else%
    \global\@minipagefalse%
    \hbox to\hsize{\box\@tempboxa}
  \fi%
  \addtolength{\hsize}{-\algomargin}%
}
\makeatother

\newcommand{\var}{\mathrm{var}}

\newcommand{\e}{E}
\newcommand{\mG}{G_m^2}
\newcommand{\tG}{G_t^2}
\newcommand{\gs}{G^2}
\newcommand{\Rs}{R^2}
\newcommand{\hD}{D}
\newcommand{\hs}{\widehat{\sigma}}
\newcommand{\hn}{\widehat{\nu}}
\newcommand{\h}{\mathcal{H}}
\newcommand{\prob}{\mathrm{pr}}
\newcommand{\ace}{\citep{Breiman:1985}}
\newcommand{\cop}{\citep{Genest:2004}}
\newcommand{\dcor}{\citep{Szekely:2007}}
\newcommand{\ddp}{\citet{Heller:2016}}
\newcommand{\ecf}{\citep{Kankainen:1998}}
\newcommand{\hoef}{\citep{Hoeffding:1948}}
\newcommand{\ik}{\citep{Kraskov:2004}}
\newcommand{\mic}{\citep{Reshef:2011}}

\DeclarePairedDelimiter{\ceil}{\lceil}{\rceil}

\begin{document}

\markboth{X. Wang, B. Jiang \and J. S. Liu}{Generalized R-squared for detecting dependence}

\title{Generalized R-squared for detecting dependence}

\author{X. WANG}
\affil{Department of Statistics, Harvard University, Cambridge, Massachusetts,  U.S.A \email{xufeiwang@fas.harvard.edu}}

\author{B. JIANG}
\affil{Two Sigma Investments, Limited Partnership, New York, New York, U.S.A  \email{bojiang83@gmail.com}}

\author{\and J. S. LIU}
\affil{Department of Statistics, Harvard University, Cambridge, Massachusetts, U.S.A \email{jliu@stat.harvard.edu}}

\maketitle
\begin{abstract}
Detecting dependence between two random variables is a fundamental problem. Although the Pearson correlation is effective for capturing linear dependency, it can be entirely powerless for detecting nonlinear and/or heteroscedastic patterns. We introduce a new measure, G-squared, to test whether two univariate random variables are independent and to measure the strength of their relationship. The G-squared is almost identical to the square of the Pearson correlation coefficient, R-squared, for linear relationships with constant error variance, and has the intuitive meaning of the piecewise R-squared between the variables. It is particularly effective in handling nonlinearity and heteroscedastic errors.  We propose two estimators of G-squared and show their consistency. Simulations demonstrate that G-squared estimates are among the most powerful test statistics compared with several state-of-the-art methods.
\end{abstract}


\section{Introduction}
\label{sec:int}
The Pearson correlation coefficient is widely used to detect and measure the dependence of two random quantities. The square of its least-squares estimate, popularly known as R-squared, is often used to quantify how linearly related two random variables are. However, the shortcomings of the R-squared as a measure of the strength of dependence are also significant, as discussed recently by \citet{Reshef:2011}, which has inspired the development of many new methods for detecting dependence.

The Spearman correlation calculates the Pearson correlation coefficient between rank statistics. Although more robust than the Pearson correlation, this method still cannot capture non-monotone relationships. The alternating conditional expectation method was introduced by \citet{Breiman:1985}  to approximate the maximal correlation between $X$ and $Y$, i.e., to find the optimal transformations of the data, $f(X)$ and $g(Y)$, so that their correlation is maximized. The implementation of the method has its limitations because it is unfeasible to search over all possible transformations. Estimating mutual information is another popular approach due to the fact that the mutual information is zero if and only if $X$ and $Y$ are independent. Furthermore, \citet{Kraskov:2004} proposed an efficient method by estimating the entropy of $X$, $Y$ and $(X,Y)$ separately. The method was claimed to be numerically exact for independent cases and to also work for high dimensional variables. An energy distance-based method \citep{Szekely:2007,Szekely:2009} and a kernel-based method \citep{Gretton:2005,Gretton:2012} appeared separately in statistics and machine learning literature to solve the two-sample test problem and have corresponding usage in independence tests. The two methods were recently shown to be equivalent \citep{Sejdinovic:2013}. Methods based on empirical cumulative distribution functions \citep{Hoeffding:1948}, empirical copula \citep{Genest:2004} and empirical characteristic functions \citep{Kankainen:1998, Huskova:2008} have also been proposed for detecting dependence.

Another set of approaches is based on discretizations of the random variables. Known as grid-based methods, they are primarily designed to test independence between univariate random variables. \citet{Reshef:2011} introduced a new statistic, the maximum information coefficient, which focuses on the generality and equitability of a dependence statistic.  Y.~Reshef and coauthors (arXiv:1505.02213) proposed two new estimators for this quantity, which are empirically more powerful and easier to compute. \citet{Heller:2016} proposed a grid based method, which utilizes the $\chi^2$ statistic to test independence and is a distribution-free test. 

To measure how accurately an independence test can reflect the strength of  dependence between two random variables,  \citet{Reshef:2011} introduced the idea of equitability, which was more carefully defined and examined in (Y.~Reshef and coauthors, arXiv:1505.02212). Equitability requires that the same value of the statistic implies the same amount of dependence, regardless of the type of relationship. Whether there exists a statistic that can achieve exact equitability is still subject to debate. However, given a collection of functional relationships with varying noise levels, we can compare the empirical equitability of different statistics through simulation studies.

Intuitively, if there is a functional relationship between two random variables $X$ and $Y$, it is natural to estimate their relationship using a nonparametric technique and use the fraction of reduction in the sum of squares as a measure of the strength of the relationship. In this way, one can both detect dependence and provide an equitable statistic. In contrast, it is more challenging for other types of dependence measures, such as energy-based or entropy-based methods, to be equitable. \citet{Doksum:1994} and \citet{Blyth:1994} discussed the correlation curve to measure the strength of the relationship. However, a direct use of nonparametric curve estimation may rely too heavily on the smoothness assumption of the relationship; it also cannot deal with heteroscedastic noises.

The $\gs$ proposed in this paper is derived from a regularized likelihood ratio test for piecewise linear relationships and can be viewed as an integration of continuous and discrete methods. The G-squared statistic is a function of both the conditional mean and conditional variance of one variable given the other. It is thus capable of detecting general functional relationships with heteroscedastic error variances. An estimate of $\gs$ can be derived via the same likelihood ratio approach as the $\Rs$ when the true underlying relationship is linear. Thus, it is reasonable that $\gs$ is almost identical to the $\Rs$ for linear relationships.
Efficient estimates of  $\gs$ can  be computed quickly by a dynamic programming method, whereas \citet{Reshef:2011}  and \cite{Heller:2016} have to consider grids on two variables simultaneously and hence require longer computational time, as shown by our simulation studies. We will also show that, in terms of both power and equitability, $\gs$ is among the best statistics for independence testing in consideration of a wide range of functional relationships.  

\section{Measuring dependence with G-squared}
\label{sec:def}
\subsection{Defining the $\gs$ as a generalization of the $\Rs$}\label{subsec:def}
The R-squared measures how well the data fit a linear regression model. Given $Y=\mu+\beta X+e$ with $e\sim N(0,\sigma^2)$, the standard estimate of R-squared can be derived from a likelihood ratio test statistic for testing $\h_0:\beta=0$ against $\h_1:\beta\neq 0$, i.e.,
\begin{eqnarray*}
\Rs = 1-\left(\frac{L(\widehat{\theta})}{L_0(\widehat{\theta}_0)}\right)^{-2/n},
\end{eqnarray*}
and $L_0(\widehat{\theta}_0)$ and $L(\widehat{\theta})$ are the maximized likelihoods under $\h_0$ and $\h_1$. 

Throughout the paper, we let $X$ and $Y$ be univariate continuous random variables. As a working model, we assume that the relationship between $X$ and $Y$ can be characterized as $Y =f(X)+\epsilon\sigma_X$, $\epsilon \sim N(0,1)$ and $\sigma_X>0$. If $X$ and $Y$ are independent, then $f(X)\equiv\mu$ and $\sigma^2_X\equiv\sigma^2$. Now, let us look at the piecewise linear relationship
\begin{eqnarray*}
f(X)=\mu_h+\beta_h X,\quad \sigma^2_X=\sigma^2_h,\quad c_{h-1}<X\leq c_h,
\end{eqnarray*}
where $c_h \ (h=0,\ldots,K)$ are  called the breakpoints. While this working model allows for heteroscedasticity, it requires constant variance within each segment between two adjacent breakpoints. Testing whether $X$ and $Y$ are independent is equivalent to testing whether $\mu_h=\mu$ and $\sigma_h^2=\sigma^2$. Given $c_h\ (h=0,\ldots,K)$, the likelihood ratio test statistic can be written as 
\begin{eqnarray*}
\textsc{lr} = \exp\left(\frac{n}{2}\log \hn^2  - \sum_{h=1}^K \frac{n_h}{2}\log \hs^2_h\right),
\end{eqnarray*}
where $\hn^2$ is the overall sample variance of $Y$ and $\hs^2_h$ is the residual variance after regressing $Y$ on $X$ for $X\in (c_{h-1}, c_h]$. Because $\Rs$ is a transformation of the likelihood ratio and converges to the square of Pearson correlation coefficient, we perform the same transformation on $\textsc{lr}$. The resulting test statistic converges to a quantity related to the conditional mean and the conditional variance of $Y$ on $X$. It is easy to show that, as $n\to\infty$, 
\begin{eqnarray}
1 - \left(\textsc{lr}\right)^{-2/n}\to 1-\exp\left[\e\{\log\var(Y\mid X)\} - \log\var(Y)\right].
\label{equ:gyx}
\end{eqnarray}
When $h=1$, the relationship degenerates to a simple linear relationship and $1 - \left(\textsc{lr}\right)^{-2/n}$ is exactly $\Rs$. 

More generally, because a piecewise linear function can approximate any almost-everywhere continuous function, we can employ the same hypothesis testing framework as above to derive  (\ref{equ:gyx}) for any such approximation. Thus, for any pair of random variables $(X,Y)$, the following concept is a natural generalization of the R-squared:
\begin{eqnarray*}
\gs_{Y\mid X} &=&1 - \exp\left[\e\{\log\var(Y\mid X)\} - \log\var(Y)\right],
\end{eqnarray*}
in which we require that $\var(Y)<\infty$. Evidently, $\gs_{Y\mid X}$ lies between zero and one, and is equal to zero if and only if both $\e(Y\mid X)$ and $\var(Y\mid X)$ are constant. The definition of $\gs_{Y\mid X}$ is closely related to the R-squared defined by segmented regression \citep{Oosterbaan:2006} discussed in the Supplementary Material. We symmetrize $\gs_{Y\mid X}$ to arrive at the following quantity as the definition of the G-squared:
\begin{eqnarray*}
\gs=\max(\gs_{Y\mid X},\  \gs_{X\mid Y}),
\end{eqnarray*}
provided $\var(X)+\var(Y)<\infty$. Thus, $\gs = 0$ if and only if $\e(X\mid Y)$, $\e(Y\mid X)$, $\var(Y\mid X)$ and $\var(X\mid Y)$ are all constant, which is not equivalent to independence of $X$ and $Y$. In practice, however, dependent cases with $\gs=0$ are rare.

\subsection{Estimation of $\gs$}
Without loss of generality, we focus on the estimation of $\gs_{Y\mid X}$; $\gs_{X\mid Y}$ can be estimated in the same way by flipping $X$ and $Y$. When $Y = f(X)+ \epsilon\sigma_X$ and $\epsilon\sim N(0,1)$ for an almost-everywhere continuous function $f(\cdot)$, we can use a piecewise linear function to approximate $f(X)$ and estimate $\gs$. However, in practice the number and locations of the breakpoints are unknown. We propose two estimators of $\gs_{Y\mid X}$, the first aiming to find the maximum penalized likelihood ratio among all possible piecewise linear approximations, and the second focusing on a Bayesian average of all approximations.

Suppose we have $n$ sorted independent observations,  $(x_i,y_i) \ (i=1,\ldots,n)$, such that $x_1<\cdots <x_n$. For the set of breakpoints, we only need to consider $c_h=x_i$. Each interval $s_h=(c_{h-1}, c_h]$  is called a slice of the observations, so that $c_h\ (h=0,\ldots,K)$ divide the range of $X$ into $K$ non-overlapping slices. Let $n_h$ denote the number of observations in slice $h$, and let $S(X)$ denote a slicing scheme of $X$, that is, $S(x_i)=h$ if $x_i \in s_h$, which is abbreviated as $S$ whenever the meaning is clear. Let $|S|$ be the number of slices in $S$ and let $m_S$ denote the minimum size of all the slices. 

To avoid overfitting when maximizing log-likelihood ratios over both unknown parameters and all possible slicing schemes, we restrict the minimum size of each slice as $m_S\geq \ceil{n^{1/2}}$ and maximize the log-likelihood ratio with a penalty on the number of slices. For simplicity, let $m=\ceil{n^{1/2}}$.  Thus,  we focus on the following penalized log-likelihood ratio
\begin{equation}
n\hD(Y\mid S,\lambda_0) = 2 \log \textsc{lr}_{S}-\lambda_0(|S|-1)\log n, \label{d-stat}
\end{equation}
where $\textsc{lr}_{S}$ is the likelihood ratio for $S$ and $\lambda_0\log n > 0$ is the penalty for incurring one additional slice.  From a Bayesian perspective, this is equivalent to assigning the prior distribution for the number of slices to be proportional to $n^{-\lambda_0(|S|-1)/2}$. Suppose each observation $x_i\ (i=1,\ldots,n-1)$ has probability $p_n=n^{-\lambda_0/2}/(1+n^{-\lambda_0/2})$ of being the breakpoint independently. Then the probability of a slicing scheme $S$ is 
\[p_n^{|S|-1}(1-p_n)^{n-|S|}\propto\left(\frac{p_n}{1-p_n}\right)^{|S|-1}= n^{-\lambda_0(|S|-1)/2}.\] 
When $\lambda_0=3$, the statistic $-n\hD(Y\mid S,\lambda_0) $ is equivalent to the Bayesian information criterion \citep{Schwarz:1978} up to a constant. 

Treating the slicing scheme as a nuisance parameter, we can maximize over all allowable slicing schemes to obtain that
\[\hD(Y\mid X,\lambda_0) = \max_{m_S\geq m} \hD(Y\mid S,\lambda_0).\]
Our first estimator of $\gs_{Y\mid X}$, which we call $\mG$ with m representing the maximum likelihood ratio, can be defined as
\begin{equation*}
\mG(Y\mid X,\lambda_0) = 1- \exp\{-\hD(Y\mid X,\lambda_0)\}.
\end{equation*}
Thus, the overall G-squared can be estimated as
\[
\mG(\lambda_0) = \max\{\mG(Y\mid X,\lambda_0),\  \mG(X\mid Y,\lambda_0)\}.
\]
By definition, $\mG(\lambda_0)$ lies between 0 and 1 and $\mG(\lambda_0)=\Rs$ when the optimal slicing schemes for both directions have only one slice. Later, we will show that when $X$ and $Y$ are a bivariate normal, $\mG(\lambda_0)=\Rs$ almost surely for large $\lambda_0$.

Another attractive way to estimate $\gs$ is to integrate out the nuisance slicing scheme parameter. A full Bayesian approach would require us to compute the Bayes factor \citep{Kass:1995}, which may be undesirable since we do not wish to impose too strong a modeling assumption. On the other hand, however, the Bayesian formalism may guide us to a desirable integration strategy for the slicing scheme. We thus put the problem into a Bayes framework and compute the Bayes factor for comparing the null and alternative models. The null model is only one model while the alternative is any piecewise linear model, possibly with countably infinite pieces. Let $p_0(y_1,\ldots,y_n)$ be the marginal probability of the data under the null. Let $\omega_{S}$  be the prior probability for slicing scheme $S$  and let $p_{S}(y_1,\ldots,y_n)$ denote the marginal probability of the data under $S$. The Bayes factor can be written as
\begin{eqnarray}
\textsc{bf}&=&\sum_{m_s\geq m}\omega_{S} \times  p_{S}(y_1,\ldots,y_n)/p_0(y_1,\ldots,y_n).
\label{equ:bf}
\end{eqnarray}
The marginal probabilities are not easy to compute even with proper priors. 
\citet{Schwarz:1978} states that if the data distribution is in the exponential family and the parameter is of dimension $k$, the marginal probability of the data can be approximated as
\begin{eqnarray}
p(y_1,\ldots,y_n) \approx \textsc{l}\exp\left\{-k(\log n -\log 2\pi)/2\right\},
\label{equ:bic}
\end{eqnarray}
where $\textsc{l}$ is the maximized likelihood. In our setup, the number of parameters $k$ for the null model is two, and for an alternative model with a slicing scheme $S$ is $3|S|$. Plugging expression (\ref{equ:bic}) into both the numerator and the denominator of (\ref{equ:bf}), we obtain
\begin{eqnarray}
\textsc{bf} \approx \sum_{S: \ m_s\geq m} \omega_{S} \textsc{lr}_{S}\exp\left\{-(3|S|-2)(\log n-\log 2\pi)/2\right\}.\label{bf-approx}
\end{eqnarray}
If we take $\omega_{S} \propto n^{-\lambda_0(|S|-1)/2}\ (\lambda_0>0)$,  which corresponds to the penalty term in (\ref{d-stat}) and is involved in defining $\mG$, the approximated Bayes factor can be restated as
\begin{equation}
\textsc{bf}(\lambda_0)=\left\{\sum_{S: \ m_S\geq m} n^{-\frac{\lambda_0(|S|-1)}{2}}\right\} ^{-1}\sum_{S:
\ m_S\geq m} \left(\frac{2\pi}{n}\right)^{\frac{3|S|-2}{2}}\exp\left\{ \frac{n}{2} D(Y\mid S,\lambda_0)\right\}.\label{bf-original} 
\end{equation}
As we will discuss in Section~\ref{lambda0}, $\textsc{bf}(\lambda_0)$ can serve as a marginal likelihood function for $\lambda_0$ and can be used to find an optimal $\lambda_0$ suitable for a particular data set. This quantity also looks like an average version of $\mG$, but with an additional penalty.  Since $\textsc{bf}(\lambda_0)$ can take values below 1, its transformation $1- \textsc{bf}(\lambda_0)^{-2/n}$, as in the case where we derived the $\Rs$ via the likelihood ratio test, can take negative values, especially when $X$ and $Y$ are independent, and it is therefore not an ideal estimator of $\gs$. 

By removing the model size penalty term in (\ref{bf-approx}), we obtain a modified version, which is simply a weighted average of the likelihood ratios and is guaranteed to be greater than or equal to 1:  
\begin{eqnarray*}
\textsc{bf}^*(\lambda_0)=\left\{\sum_{S: \ m_S\geq m} n^{-\frac{\lambda_0(|S|-1)}{2}} \right\} ^{-1}\sum_{S: m_S\geq m} \exp\left\{\frac{n}{2}\hD(Y\mid S,\lambda_0)\right\}.
\end{eqnarray*}
We can thus define a quantity similar to our likelihood formulation of R-squared,
\[\tG(Y\mid X,\lambda_0)=1-\textsc{bf}^*(\lambda_0)^{-2/n},\]
which we call the total G-squared, and define 
\[ \tG(\lambda_0)=\max\{\tG(Y\mid X,\lambda_0),\  \tG(X\mid Y,\lambda_0)\}.\] 
We show later that $\mG(\lambda_0)$  and $\tG(\lambda_0)$ are both consistent estimators of $\gs$. 

\subsection{Theoretical properties of the $\gs$ estimators}
\label{sec:pro}
In order to show that $\mG(\lambda_0)$ and $\tG(\lambda_0)$ converge to  $\gs$ as the sample size goes to infinity, we introduce the notations: $\mu_X(y) = \e(X\mid Y=y)$,  $\mu_Y(x) = \e(Y\mid X=x)$,  $\nu_X^2(y) = \var(X\mid Y=y)$ and $\nu_Y^2(x) = \var(Y\mid X=x)$ as well as the following regularity conditions:

\begin{condition}
The random variables $X$ and $Y$ are bounded continuously with finite variances such that $\nu_Y^2(x)$, $\nu_X^2(y)>b^{-2}>0$ almost everywhere for some constant $b$.
\end{condition}
\begin{condition}
The functions $\mu_Y(x)$, $\mu_X(y)$, $\nu_Y^2(x)$ and $\nu_X^2(y)$ have continuous derivatives almost everywhere.
\end{condition}
\begin{condition}
There exists a constant $C > 0$ such that
\begin{eqnarray*}
\max\{\left|\mu_X'(y)\right|,\ \left|\nu_X'(y)\right|\} \leq C\nu_X(y),\quad \max\{\left|\mu_Y'(x)\right|,\ \left|\nu_Y'(x)\right|\} \leq C\nu_Y(x)
\end{eqnarray*}
almost surely.
\end{condition}

With these preparations, we can state our main results.
\begin{theorem}
\label{the:1} Under Conditions 1-3, for all $\lambda_0 > 0$, 
\begin{eqnarray*}
\mG(Y\mid X,\lambda_0)\rightarrow \gs_{Y\mid X}, \quad \tG(Y\mid X,\lambda_0)\rightarrow \gs_{Y\mid X}
\end{eqnarray*}
almost surely as $n\to\infty$. Thus, $\mG(\lambda_0)$ and $\tG(\lambda_0)$ are consistent estimators of $\gs$. 
\end{theorem}

A proof of the theorem and numerical studies of the consistency are in the Supplementary Material. It is expected that $\mG(\lambda_0)$ should converge to $\gs$ just because of its construction. It is surprising that $\tG(\lambda_0)$ also converges to $\gs$. The result, which links  $\gs$ estimation with the likelihood ratio and Bayesian formalism, suggests that most of the information up to the second moment has been fully utilized in the two test statistics. The  theorem thus supports the use of $\mG(\lambda_0)$ and $\tG(\lambda_0)$ for testing whether $X$ and $Y$ are independent. The null distributions of the two statistics depend on the marginal distributions of $X$ and $Y$, which can be generated empirically using  permutation. One can also do a quantile-based transformation on $X$ and $Y$ such that their marginal distributions are standard normal; however, the $\gs$ based on the transformed data tends to lose some power.

When $X$ and $Y$ are bivariate normal, the G-squared statistic is almost the same as the R-squared when $\lambda_0$ is large enough.
\begin{theorem}
\label{the:2}
If $X$ and $Y$ follow bivariate normal distribution, then for $n$ large enough
\begin{eqnarray*}
\prob \left\{\mG(\lambda_0) = \Rs\right\} &>& 1-3n^{-\lambda_0/3+5}.
\end{eqnarray*}
So for $\lambda_0>18$ and $n\to\infty$, we have $\mG(\lambda_0)=\Rs $ almost surely .
\end{theorem}
The lower bound on $\lambda_0$ is not tight and can be relaxed in practice. Empirically, we have observed that $\lambda_0=3$ is large enough for $\mG(\lambda_0)$ to be very close to $\Rs$ in the bivariate normal setting.

\subsection{Dynamic programming algorithm for computing $\mG$ and $\tG$}\label{dp}
The brute force calculation of either $\mG$ or $\tG$ has a computational complexity of $O(2^n)$ and is prohibitive in practice. Fortunately, we have found a dynamic programming scheme for computing both quantities with a time complexity of $O(n^2)$.  The algorithms for computing  $\mG(Y\mid X,\lambda_0)$ and $\tG(Y\mid X,\lambda_0)$ are roughly the same except for one operation, i.e., maximization versus summation, and can be summarized by the following steps:

\begin{step}[Data preparation] Arrange the observed pairs $(x_i, y_i)\ (i=1,\ldots,n)$ according to the sorted $x$s from low to high. Then normalize $y_i \ (i=1,\ldots,n)$ such that $\sum_{i=1}^ny_i=0$ and $\sum_{i=1}^n y_i^2=1$.
\end{step}
\begin{step}[Main algorithm] Define $m=\ceil{n^{1/2}}$ as the smallest slice size, $\lambda=-\lambda_0\log(n)/2$ and $\alpha=e^\lambda$. Initialize three sequences: $(M_i,\ B_i,\ T_i) \  (i=1,\ldots,n)$ with $M_1=0$ and $B_1=T_1=1$. For $i=m,\ldots, n$, recursively fill in entries of the tables with
\begin{eqnarray*}
M_i = \max_{k\in K_i}\left( \lambda + M_k + l_{k:i}\right),\quad B_i = \sum_{k\in K_i} \alpha B_k, \quad T_i = \sum_{k\in K_i} \alpha T_k L_{k:i},
\end{eqnarray*}
where $K_i=\{1\}\cup\{k:k=m+1,\ldots, i-m+1\}$, $l_{k:i} = -(i-k)\log(\hs_{k:i}^2)/2$ and $L_{k:i}=\exp\{l_{k:i}\}$, with $\hs_{k:i}^2$ as the residual variance of regressing $y$ on $x$ for observations $(x_{j}, y_{j}) \ (j=k,\ldots,i)$.
\end{step}
\begin{step} The final result is
\begin{eqnarray*}
\mG = 1-\exp\left\{ M_n - \lambda \right\},\quad \tG = 1- (T_n/B_n)^{-2/n}.
\end{eqnarray*}
\end{step}
Here, $M_i\ (i=m,\ldots, n)$ stores the partial maximized likelihood ratio up to the ordered observation $(x_k,y_k) \ (k=1,\ldots, i)$, $B_i \ (i=m,\ldots, n)$ stores the partial normalizing constant, and $T_i \ (i=m,\ldots, n)$ stores the partial sum of the likelihood ratios. When $n$ is extremely large, we can speed up the algorithm by considering fewer slice schemes. For example, we can divide $X$ into chunks of size $m$ by rank and consider only slicing schemes between the chunks. For this method, the computational complexity is $O(n)$. We can compute $\mG(X\mid Y,\lambda_0)$ and $\tG(X\mid Y,\lambda_0)$ similarly to get $\mG(\lambda_0)$ and $\tG(\lambda_0)$. Empirically, the algorithm is faster than many other powerful methods as shown in the Supplementary Material.

\subsection{An empirical Bayes strategy for selecting $\lambda_0$} 
\label{lambda0}

\begin{figure}
\centering{\includegraphics[height=2.4in,width=4.8in]{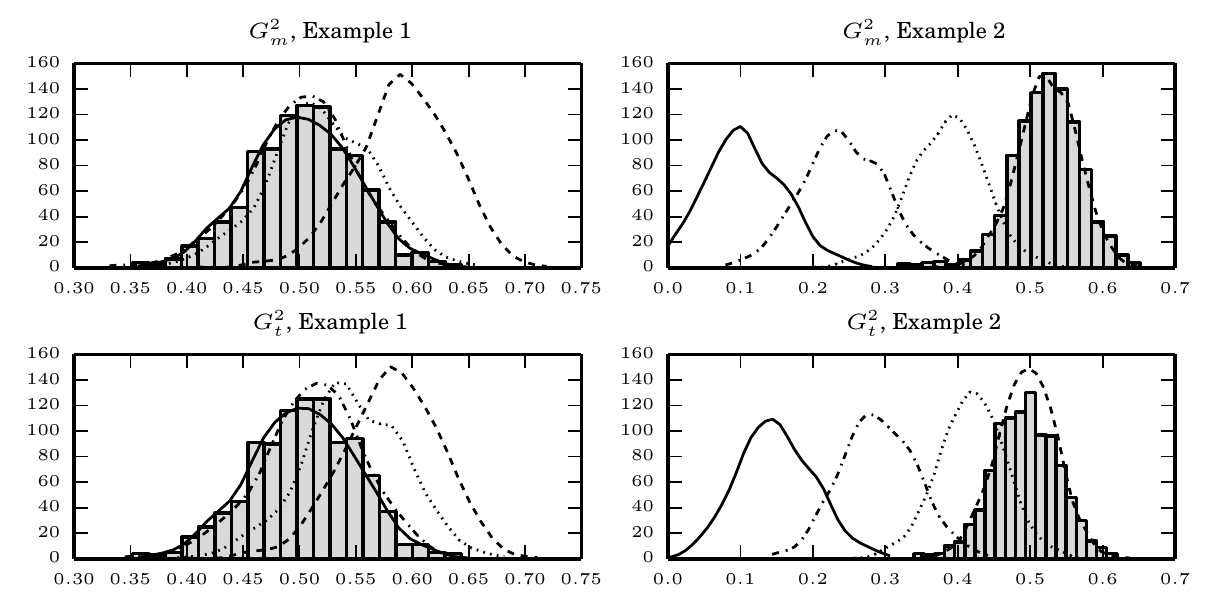}}
\caption{Sampling distributions of $\mG$ and $\tG$ under the two models in Section~\ref{lambda0} with $\gs_{Y\mid X}=0.5$ for $\lambda_0$ = 0.5 (dashes), 1.5 (dots), 2.5 (dot-dash) and 3.5 (solid). The density function in each case is estimated by the histogram. The sampling distributions of $\mG$ and $\tG$ with the empirical Bayes selection of $\lambda_0$  are in gray shadow and overlaid on top of other density functions.}
\label{fig:den_0.5_part}
\end{figure}

Although the choice of the penalty parameter $\lambda_0$ is not critical for the general use of $\gs$, we typically use $\lambda_0 = 3$ for $\mG$ and $\tG$ because $\hD(Y\mid X, 3)$ is equivalent to the Bayesian information criterion. Fine-tuning $\lambda_0$ can improve the estimation of $\gs$. We thus propose a data-driven strategy for choosing $\lambda_0$ adaptively. $\textsc{bf}(\lambda_0)$ in (\ref{bf-original}) can be viewed as an approximation to $\prob(y_1,\ldots,y_n\mid \lambda_0)$ up to a normalizing constant. We thus can use the maximum likelihood principle to choose the $\lambda_0$ that maximizes $\textsc{bf}(\lambda_0)$. We then use the chosen $\lambda_0$  to  compute $\mG$ and $\tG$ as estimators of $\gs$. In practice, we evaluate $\textsc{bf}(\lambda_0)$ for a set of discrete $\lambda_0$ values, e.g., $\{0.5,\ 1,\ 1.5,\ 2,\ 2.5,\ 3,\ 3.5,\ 4\}$, and pick the one that maximizes $\textsc{bf}(\lambda_0)$. $\textsc{bf}(\lambda_0)$ can be computed efficiently via a dynamic programming algorithm similar to that described in Section~\ref{dp}. As an illustration, we consider the sampling distributions of  $\mG(\lambda_0)$ and $\tG(\lambda_0)$ with $\lambda_0=0.5,\ 1.5,\ 2.5$ and $3.5$ for the following two examples
\begin{example} $X\sim N(0,1)$, $Y=X+\sigma \epsilon$ and $\epsilon\sim N(0,1)$.
\end{example} 
\begin{example}
$X\sim N(0,1)$, $Y=\sin(4\pi x)/0.7+\sigma \epsilon $ and $\epsilon\sim N(0,1)$.
\end{example}
We simulated $n=225$ data points. For each model, we set $\sigma=1$ so that $\gs_{Y\mid X}=0.5$ and performed 1,000 replications. Figure~\ref{fig:den_0.5_part} shows histograms of $\mG(\lambda_0)$ and $\tG(\lambda_0)$ with different $\lambda_0$ values. The results demonstrate that, for relationships that can be approximated well by a linear function, a larger $\lambda_0$ is preferred because it penalizes the number of slices more heavily and the resulting sampling distributions are less biased. On the other hand, for complicated relationships such as the trigonometric function, a smaller $\lambda_0$ is preferable because it allows more slices, which can help capture fluctuations in the functional relationship. The figure also shows that the empirical Bayes selection of $\lambda_0$ worked very well, leading to a proper choice of $\lambda_0$ for each simulated data set from both examples and resulting in the most accurate estimates of $\gs$. Additional simulation studies and consistency of the data-driven strategy are in the Supplementary Material.

\section{Simulation Studies}
\label{sec:sim}
\subsection{Power analysis}
\label{subsec:pow}
Now we compare the power of different independence testing methods for various types of relationships. Here, we again fixed $\lambda_0=3$ for both $\mG$ and $\tG$. Other methods we tested include the alternating conditional expectation \ace, Genest's test \cop, Pearson correlation, distance correlation \dcor, the method of \ddp, the characteristic function method \ecf, Hoeffding's test \hoef, the mutual information method \ik \ and two methods, $\textsc{mic}_e$ and $\textsc{tic}_e$, based on the maximum information criterion \mic. We follow the procedure for computing the powers of different methods as described in previous studies
of (D.~Reshef and coauthors, arXiv:1505.02214) and a 2012 online note by N.~Simon and R.~Tibshirani.

For different functional relationships $f(X)$ and different values of noise levels $\sigma^2$, we simulated $(X,Y)$ with the following model:
\begin{eqnarray*}
X\sim U(0,1),\quad Y=f(X)+\epsilon \sigma,\quad \epsilon \sim N(0,1).
\end{eqnarray*}
where $\var\{f(X)\}=1$. Thus $\gs_{Y\mid X}=(1+\sigma^2)^{-1}$ is a monotone function of the signal-to-noise ratio and it is of interest for us to observe how the performances of different methods deteriorate as the signal strength weakens for various functional relationships. We used permutations to generate the null distribution and to set the rejection region for all testing methods in all examples. 

\begin{figure}
\centering{\includegraphics[height=5.12in, width=4.8in]{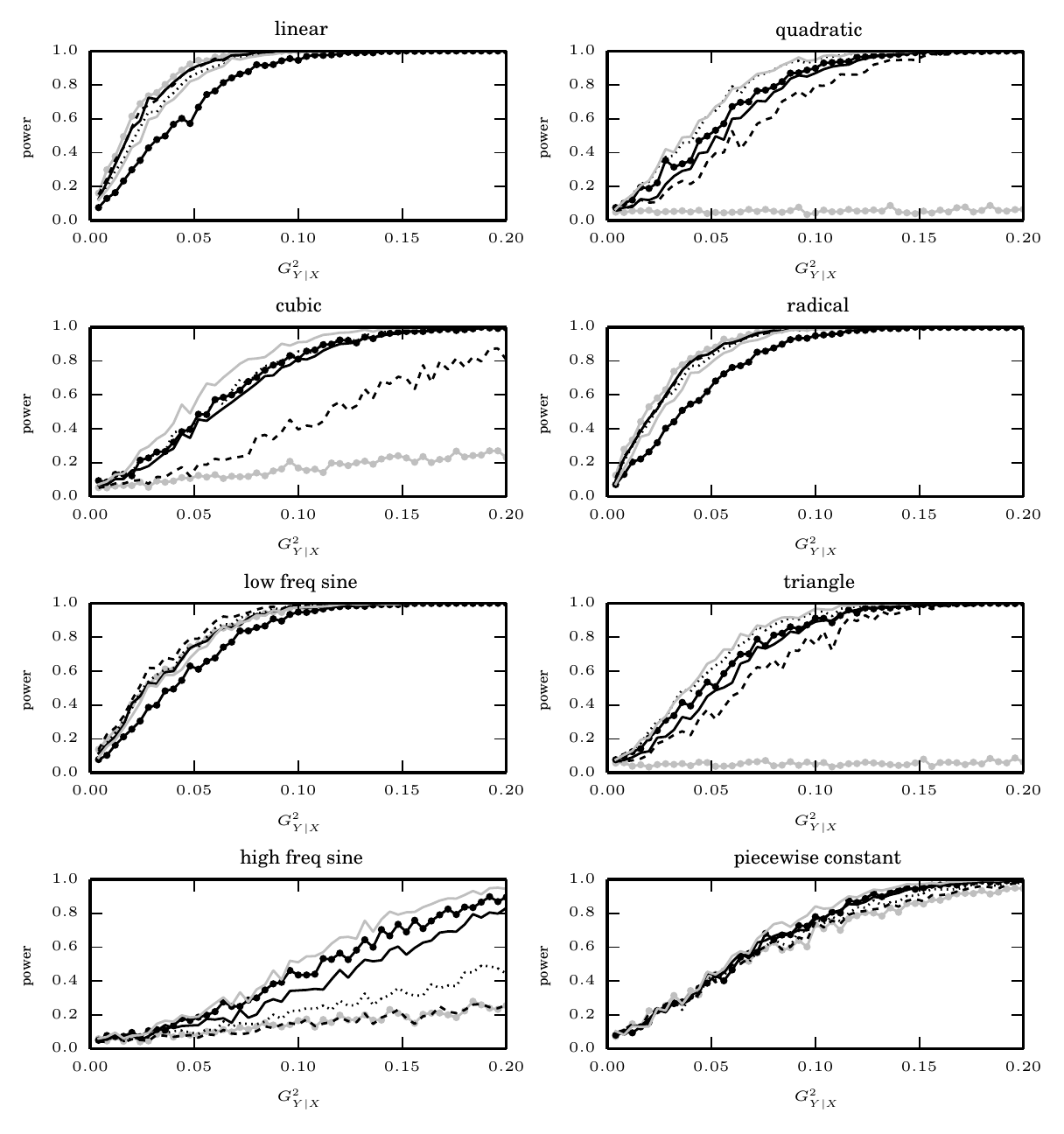}}
\caption{The powers of $\mG$ (black solid), $\tG$ (grey solid), Pearson correlation (grey markers), distance correlation (black dashes), method of \ddp \ (black dots) and $\textsc{tic}_e$ (black markers) for testing independence  between $X$ and $Y$ when the underlying true functional relationships are linear, quadratic, cubic, radical, low freq sine, triangle, high freq sine and piecewise constant, respectively. The x-axis is $\gs_{Y \mid X}$, a monotone function of the signal-to-noise ratio, and the y-axis is the power. We chose $n=225$ and performed 1,000 replications for each relationship and $\gs_{Y\mid X}$.}
\label{fig:pow}
\end{figure}

Figure~\ref{fig:pow} shows the power comparisons for eight  functional relationships. We set the sample size $n=225$ and performed 1,000 replications for each relationship and $\gs_{Y\mid X}$ value. We only plot Pearson correlation, distance correlation, method by \ddp, $\textsc{tic}_e$, $\mG$ and $\tG$ for a clear presentation. More simulations are in the Supplementary Material. For any method with tuning parameters, we chose the ones that resulted in the highest average power over all the examples. Due to computational concerns, we chose $K=3$ for the method of \ddp. It is seen that $\mG$ and $\tG$ performed  robustly, always being among the most powerful methods, with $\tG$ slightly more powerful than $\mG$ in almost all examples. They outperformed other methods in cases such as the high frequency sine, triangle and piecewise constant functions, where piecewise linear approximation is more appropriate than other approaches.  For monotonic examples such as linear and radical relationships, $\mG$ and $\tG$ had slightly lower powers than Pearson correlation, distance correlation  and the method of \ddp, but were still highly competitive. 

We also studied the performances of these methods with different sample sizes, i.e. for $n=$50, 100 and 400, respectively, and found that $\mG$ and $\tG$ still showed high power regardless of $n$ although their advantages were much less obvious when $n$ is small.  More details can be found in the Supplementary Material. 
    
\subsection{Equitability}
\label{subsec:equ}
Y.~Reshef and coauthors (arXiv:1505.02212) gave two equivalent theoretical definitions of the equitability of a statistic that measures dependence. Intuitively, equitable statistics can be used to gauge the degree of dependence. They used $\Psi=\mathrm{cor}^2\{Y,f(X)\}$ to define the degree of dependence when the dependence of $Y$ on $X$ can be described by a functional relationship. When $\var(Y\mid X)$ is a constant, $\Psi\equiv \gs_{Y\mid X}$. For a perfectly equitable statistic, its sampling distribution should be almost identical for different relationships with the same $\Psi$.

\begin{figure}
\centering{\includegraphics[height=3.2in,width=4.8in]{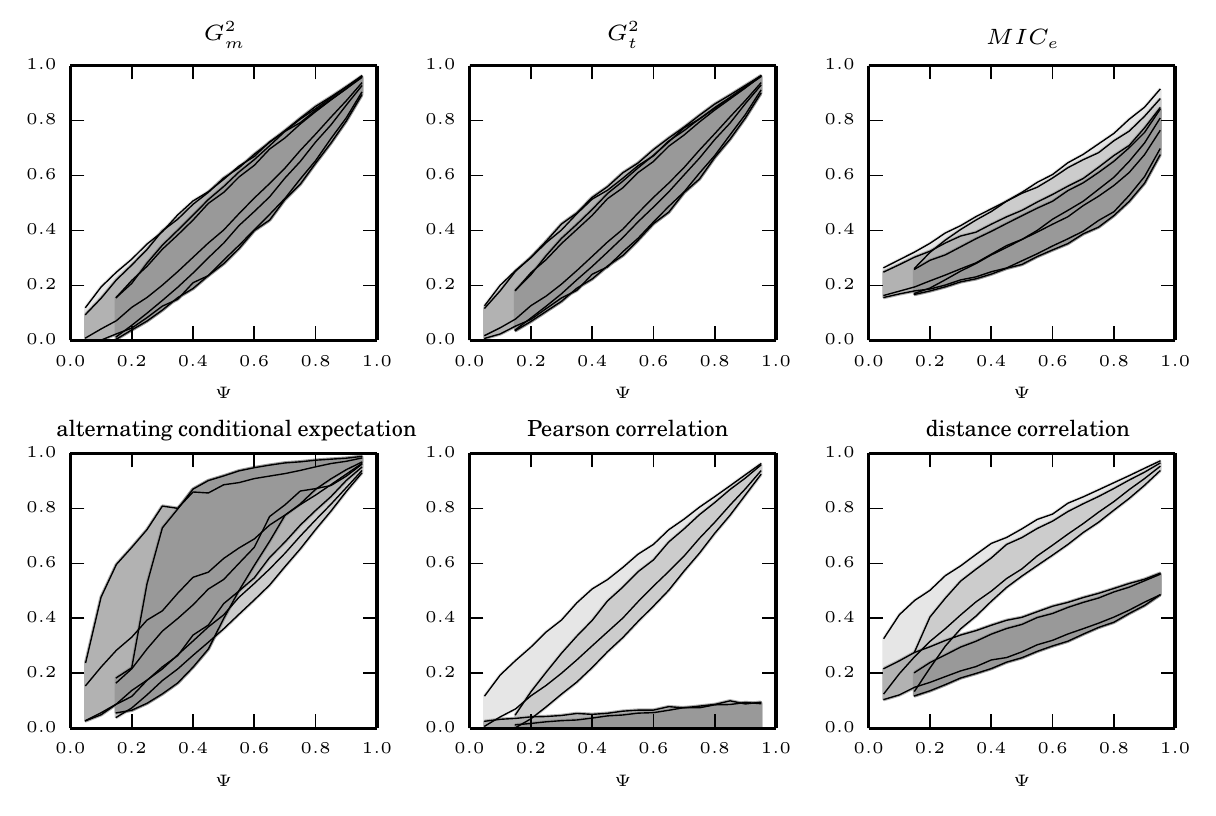}}
\caption{The plots from the top left to the bottom right are the $95\%$ confidence bands for $\mG$, $\tG$, $\textsc{mic}_e$, alternating conditional expectation, Pearson correlation and distance correlation, respectively. We chose $n=225$ and performed 1,000 replications for each relationship and each $\Psi$ value for the four examples in Section~\ref{subsec:equ}. The fill is the lightest for Example~\ref{ex:3} and darkest for Example~\ref{ex:6}.  $\Psi$ is a monotone function of the signal-to-noise ratio when the error variance is constant, and the y-axis shows the values of the corresponding statistic, each estimating its own population mean, which may or may not be $\Psi$.}
\label{fig:ei}
\end{figure}

We repeated the equitability study by \citet{Reshef:2011}. In Fig.~\ref{fig:ei}, we plot the $95\%$ confidence bands of $\mG$ and $\tG$, compared with alternating conditional expectation, Pearson correlation, distance correlation and  $\textsc{mic}_e$, for the following relationships:
\begin{example}
$X\sim U(0,1)$, $Y=X+\epsilon\sigma$ and $\epsilon\sim N(0,1)$.
\label{ex:3}
\end{example}
\begin{example}
$X\sim U(0,1)$, $Y=X+\epsilon\sigma$ and $\epsilon\sim N(0,e^{-|X|})$.
\label{ex:4}
\end{example}
\begin{example}
$X\sim U(0,1)$, $Y=\frac{X^2}{\surd{2}}+\epsilon\sigma$ and $\epsilon\sim N(0,1)$.
\label{ex:5}
\end{example}
\begin{example}
$X\sim U(0,1)$, $Y=\frac{X^2}{\surd{2}}+\epsilon\sigma $ and $\epsilon\sim N(0,e^{-|X|})$.
\label{ex:6}
\end{example}
We choose different $\Psi$ values with $n=225$ and 1,000 replications for each case. The plots show that $\mG$ and $\tG$ increased along with $\Psi$ for all relationships, as they should, and that the confidence bands obtained under different functional relationships have a similar size and location for the same $\Psi$. The confidence bands are also comparably narrow. The $\textsc{mic}_e$ displayed good performances of equitability, though slightly worse than $\mG$ and $\tG$, while other three statistics did poorly for non-monotone relationships. The alternating conditional expectation tended to have a wider confidence band for Example~\ref{ex:5} and~\ref{ex:6} than the aforementioned three methods, while Pearson correlation  and distance correlation had non-overlapping confidence intervals for different relationships when $\Psi$ is moderately large. In other words, Pearson correlation and distance correlation can yield drastically different values for two relationships with the same  $\Psi$. This phenomenon is as expected since it is known that these two statistics do not perform well for non-monotone relationships. 

\begin{figure}
\centering{\includegraphics[height=6.4in,width=4.8in]{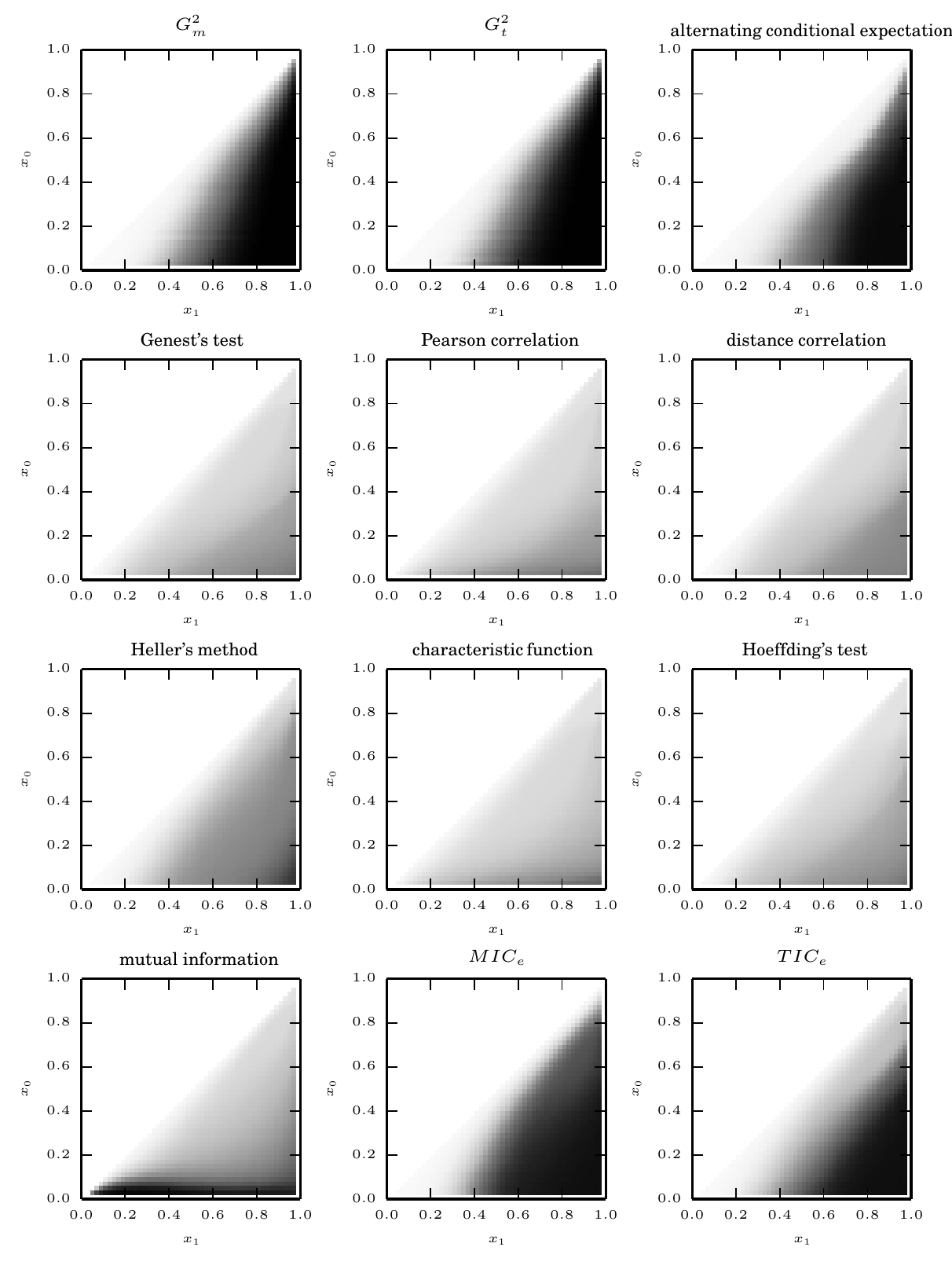}}
\caption{Heat map plot for comparing equitability of different methods. From the top left to the bottom right:  $\mG$, $\tG$, alternating conditional expectation, Genest's test, Pearson correlation, distance correlation, the method of \ddp, characteristic function method, Hoeffding's test, mutual information, $\textsc{mic}_e$ and $\textsc{tic}_e$. The value corresponding to $(x_1, x_0)\ (0<x_0<x_1<1)$ is the power of the method for testing the hypothesis: $\h_0: \Psi = x_0$ against $\h_1:\Psi = x_1$, averaging over a class of functions. The darker a dot, the higher the average power of the corresponding test. We chose sample size $n=225$ and performed 1,000 replications for each relationship and $(x_0, x_1)\ (0<x_0<x_1<1)$.}
\label{fig:equ}
\end{figure}

An alternative strategy to study equitability uses a hypothesis testing framework, i.e., to test $\h_0: \Psi  = x_0$ against $\h_1:\Psi = x_1 \ (x_1>x_0)$ on a broad set of functional relationships using a statistic. The more powerful a test statistic for this testing problem with all types of relationships, the better its equitability. For each aforementioned method, we performed  right-tailed tests with the type-I error fixed at $\alpha=0.05$ and different combinations of $(x_0, x_1)\ (0<x_0<x_1<1)$. Given a fixed sample size, a perfectly equitable statistic should yield the same power for all kinds of relationships so that it is able to reflect the degree of dependency by a single value regardless of the type of relationship. In reality, most statistics can perform well only for a small class of relationships.
In Fig.~\ref{fig:equ}, we use a heat map to demonstrate the average power of a test statistic with different pairs of $(x_0,x_1)\ (0<x_0<x_1<1)$. Each dot in the plot represents the average power of a testing method over a class of functional relationships; the darker the color is, the higher the power. We used the same set of functional relationships as in N.~Reshef and coauthors (arXiv:1505.02214) and carried out the testing for  $(x_0,x_1)=(i/50,j/50)\ (i<j=1,\ldots,49)$. We set the sample size as $n=225$ and conducted 1,000 replications for each relationship and each $(x_0, x_1)\ (0<x_0<x_1<1)$. For any method with a tuning parameter, we chose parameters that resulted in the greatest average power. We observed that $\mG$, $\tG$ and $\textsc{mic}_e$ had the best equitability, followed by alternating conditional expectation and $\textsc{tic}_e$. The average powers for $\mG$, $\tG$ and $\textsc{mic}_e$ over the entire range of $(x_0,x_1)\ (0<x_0<x_1<1)$ were all $0.6$, although $\mG$ and $\tG$ were slightly better for larger $x_0$'s.  Besides, with our empirical Bayes method for selecting $\lambda_0$, the equitability of $\mG$ and $\tG$ can be further improved. In comparison, all the  remaining methods were not as equitable. 

\section{Discussion}
The G-squared can be viewed as a direct generalization of the R-squared. While maintaining the same interpretability as the R-squared, the G-squared is also a powerful and equitable measure of dependence for general relationships. Instead of resorting to curve-fitting methods for estimating the underlying relationship and the G-squared, we employed the more flexible piecewise linear approximations with penalty and dynamic programming algorithms. Although we only consider piecewise linear functions, one can potentially approximate a relationship between two variables with piecewise polynomials or other flexible basis functions, with perhaps additional penalty terms to control the complexity.  Furthermore, it is a worthwhile effort to generalize the slicing idea for testing dependence between two multivariate random variables. 

\section*{Acknowledgment}
We are grateful to the two referees for helpful comments and suggestions. This research was supported in part by grants from the U.S. National Science Foundation and National Institutes of Health. We thank Ashley Wang for her proofreading of the paper. The views expressed herein are the authors alone and are not necessarily the views of Two Sigma Investments, Limited Partnership or any of its affiliates.

\section*{Supplementary material}
Further material available at \textit{Biometrika} online includes proofs of theorems, software implementation details, discussions on segmented regression and more simulation results.

\end{document}